# Formation of a transient amorphous solid in low density aqueous charged sphere suspensions

Ran Niu,[1,*] Sabrina Heidt,[1,2] Ramsia Sreij,[3] Riande I. Dekker,[4] Maximilian Hofmann,[1] and Thomas Palberg[1]

[1] Institute of Physics, Johannes Gutenberg University, D-55099 Mainz, Germany

[2] Graduate School Materials Science in Mainz, Staudinger Weg 9, D-55128 Mainz, Germany

[3] Department of Chemistry Physical and Biophysical Chemistry (PC III), Bielefeld University, D-33615 Bielefeld, Germany

[4] Debye Institute for Nanomaterials Science, Utrecht University, NL-3584 CC Utrecht, The Netherlands

**Correspondence to ranniu@uni-mainz.de**

**Colloidal glasses formed from hard spheres, nearly hard spheres, ellipsoids and platelets or their attractive variants, have been studied in great detail. Complementing and constraining theoretical approaches and simulations, the many different types of model systems have significantly advanced our understanding of the glass transition in general. Despite their early prediction, however, no experimental charged sphere glasses have been found at low density, where the competing process of crystallization prevails. We here report the formation of a transient amorphous solid formed from charged polymer spheres suspended in thoroughly deionized water at volume fractions of 0.0002 – 0.01. From optical experiments, we observe the presence of short-range order and an enhanced shear rigidity as compared to the stable polycrystalline solid of body centred cubic structure. On a density dependent time scale of hours to days, the amorphous solid transforms into this stable structure. We further present preliminary dynamic light scattering data showing the evolution of a second slow relaxation process possibly pointing to a dynamic heterogeneity known from other colloidal glasses and gels. We compare our findings to the predicted phase behaviour of charged sphere suspensions and discuss possible mechanisms for the formation of this peculiar type of colloidal glass.**

Glasses are amorphous solids exhibiting only short-range order and a finite shear rigidity. In atomic and molecular systems, they form from the meta-stable melt after rapid quenches in temperature or from vapour deposition. Very different substances, oxides, metals, organic molecules are able to form glasses. Glassy states of matter are also known in colloidal suspensions comprising of small particles suspended in a carrier liquid. Unlike in their atomic and molecular counterparts, however, colloidal suspensions are readily accessed by complementary optical experiments in real[1] and reciprocal space[2] as well as by theory and simulation. They are widely accepted as model systems for condensed matter problems, due to their convenient time and length scales and their analytically tractable (mostly spherically symmetric) interactions without symmetry breaking electronic degrees of freedom. Also in colloidal systems, different kinds of glasses and different routes into the glass exist. A vast body of experimental literature has been published, as discussed extensively in a number of recent excellent reviews.[3-9] Different theoretical approaches compete in elucidating the location and the mechanisms involved in forming (colloidal) glasses.[10-17] Most experiments on soft matter systems so far focused on colloidal hard spheres, which solely interact via excluded volume interactions.

These show a face centred cubic crystalline phase above their freezing transition located at a packing fraction of $\Phi_F = 0.495$ and a transition to an amorphous solid state at elevated packing fractions around $\Phi_G \approx 0.575$.[18,19] These systems greatly contributed to our current understanding of the general glass transition phenomenology and allowed checking the ranges of validity for the different theoretical approaches.[20-22] Studies on their close cousins of soft spheres, ellipsoids, platelets as well as the depletion attractive variants thereof demonstrated the influence of interaction potential shape,[23-37] particle anisotropy[28] and confinement.[29,30] Despite the wealth of experimental studies, however, it is probably fair to state, that the issue of "the" correct theory of the glass-transition is not settled and many important aspects remain to be understood. A particularly interesting point is the competition with crystallization.[31-35]

This aspect should be very pronounced in charged sphere colloids in aqueous suspension,[36,37] where jamming and the influence of gravity[38] are easily avoided, and the influence of polydispersity is much less pronounced.[39] Also these are readily accessed by optical experiments.[40] However, experimental reports on the transition of charged spheres or charged platelets into an amorphous solid or "Wigner glass"[24-26,41-46] are much rarer than theoretical studies or simulations on these systems.[47-60] In systems of charged clay platelets,[25,61-63] also the glass transition at low packing fractions was addressed and demonstrated for sufficiently strong Coulomb-couplings. There, also the occurrence of ageing processes and the existence of a second slow relaxation process was demonstrated and interpreted in terms of dynamic heterogeneities known from previous studies on hard spheres glasses. Systematic experimental studies of low density charged *sphere* glasses are so far missing even though predicted by Mode coupling Theory (MCT) approaches.[49,51-54,60]

In this work, we investigate low density amorphous solids formed from aqueous suspensions of charged spheres. Note that we intentionally denote the short-range ordered solid state of the present system as "amorphous solid" rather than as "glass", because we feel that i) many different colloidal glasses seem to co-exist in different experimental and theoretical approaches; ii) the issue of how exactly a glass can be unequivocally defined beyond being an amorphous solid therefore remains unsolved; and iii) we are not sure about the exact nature and the ways in and out of this state. Our amorphous solids are formed from standard polymer latex spheres (diameter $2a_h = 117.6$ nm, size polydispersity index PI = 0.05 (Fig. S1 in the supporting information), effective charge number $Z_{eff,G} = 379 \pm 10$) suspended in water at thoroughly deionized conditions yielding a residual ion concentration of 0.2 µmol/L. All solidification experiments were started from an initial homogenized meta-stable melt phase which was obtained mechanically by gently shaking the samples.[64] Dozens of such perfectly ordinary one component systems have previously been investigated in solidification experiments,[24,36,37,45,64-70] and all were found to exhibit an undisturbed freezing transition into body centred cubic (bcc) crystals at low to moderate particle concentrations. So far, amorphous charged sphere solids were observed only at elevated packing fraction of $\Phi \approx 0.3$-$0.5$ far above the freezing concentration.[24,26,41-44,46] Only very recently, an exception to this standard behaviour has been found.[33] We here investigate the properties of that particular system in more detail, varying the particle number density at thoroughly deionized conditions and characterizing the temporal evolution. We use static light scattering, dynamic light scattering and torsional resonance spectroscopy combined in a multi-purpose light scattering instrument[40] without the need to transfer the fragile samples from set-up to set-up. Interestingly, the amorphous solid forms at number densities of $0.2$ µm$^{-3} \leq n \leq 12$ µm$^{-3}$ corresponding to packing fractions of $1.9 \times 10^{-4} \leq \Phi \leq 0.01$.

Although it is formed at such low particle concentrations, this system nevertheless bears some strong resemblance to hard sphere or charged sphere glasses formed at elevated packing fractions. It shows a liquid-like static structure factor and at the same time displays a finite shear rigidity. From these observations, we conclude that the samples form amorphous solids. The shear modulus is systematically larger for the amorphous solids than that of the polycrystalline solid. From a comparison of this finding to theoretical expectations, we take it as an indication of a short-range order with bcc symmetry. In addition, conventional dynamic light scattering reveals that the samples develop a two-step decay in the intensity autocorrelation function indicating the emergence of a second, slow relaxation process. The nature of the second relaxation process cannot yet be resolved for this type of amorphous solid with the available equipment. Qualitatively speaking, however, a similar and presumably related phenomenology is known also from other colloidal glasses.

Our amorphous solids also show a number of peculiar features. For one, they form at very low volume fractions qualitatively compatible with the predictions of several MCT calculations based on different spherically symmetric interaction potentials.[52,60] More strikingly, the amorphous solids form at or at least very close to the freezing transition located approximately at $n_F \approx 0.15$ µm$^{-3}$. Here the equilibrium crystal phase is body centred cubic (bcc). As evidenced by shear modulus measurements, the short-range order of the amorphous solid also is of bcc symmetry. This is at difference to hard spheres which crystallize into close packed structures. In those glasses, the short-range order often is face centred cubic (fcc)-like, but sometimes it displays an icosahedral or related locally preferred structure.[2,34,35] The present amorphous solids show a pronounced competition with crystallization which always wins on long time scales. Our system of charged spheres therefore differs from systems of charged platelets, which also show an amorphous solid phase adjacent to the equilibrium fluid, but where a competing crystal phase is absent.[25,55-57] Interestingly, the kinetics of solidification change drastically and in an unexpected way with increasing packing fraction. At the lowest densities corresponding to very shallow quenches into the meta-stable melt, an amorphous solid forms within several hours, and crystallization takes days. Both time scales shorten with increasing number density. Crystallization, however accelerates much more rapidly than vitrification. Consequently, for $n > 12$ µm$^{-3}$, samples crystallize too quickly to be still unambiguously identified as an amorphous solid. Such an outcome of the competition between crystallization and vitrification has not been reported before and, in fact, is exactly opposite to that typically found in hard or charged sphere systems, where crystallization wins at low densities but is slowed and finally suppressed by vitrification at large densities.[79]

It is tempting to ascribe these distinct features to the strength and long-rangedness of the interactions of thoroughly deionized charged sphere suspensions which they share with charged plasmas[71] and so-called electron glasses.[72,73] Also for the latter systems, glassy phenomenology persists even at weak quenched disorder and is thought to be an intrinsic property of the interacting electrons themselves. The screened Coulomb repulsion between charged sphere colloids can be tailored in strength and range by three independently tuneable experimental parameters: charge, particle and electrolyte concentration.[74,75] Still, it can be conveniently parameterized by facile experiments like static light scattering, conductivity or shear modulus measurements and at the same time is well accessible for computer simulations. This opens a perspective for a large number of interesting optical studies addressing the kinetics and mechanisms of the way into and out of their amorphous state. Further, systematic variation of experimental boundary conditions will be very important for the next generation of theoretical investigations on low density charged sphere glasses.

## Results
### Static light scattering

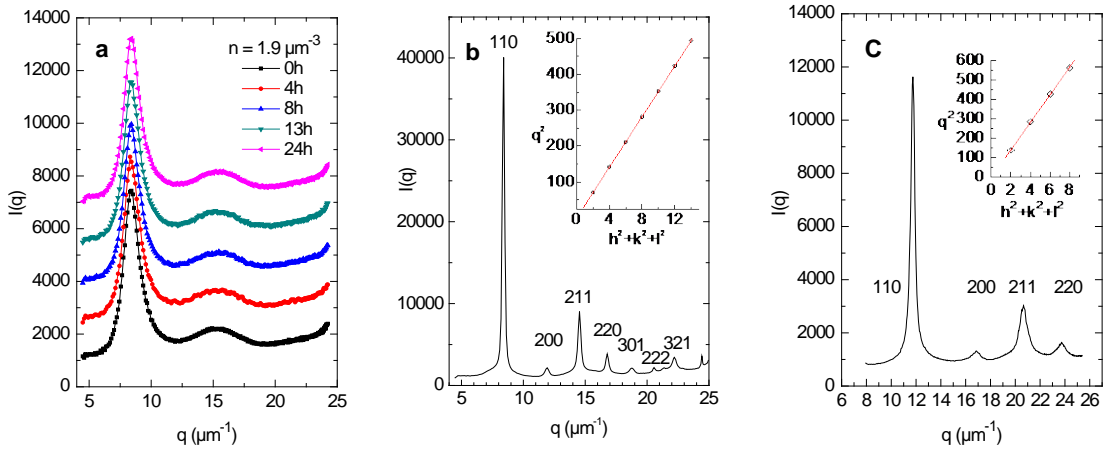

**Figure 1. Exemplary static light scattering patterns as a function of scattering vector $q$.** a) Thoroughly deionized sample of number density $n = 1.9$ µm$^{-3}$ for different times after last gentle shaking as indicated. Curves are shifted for clarity. The scattering pattern is fluid-like. b) After 3d, this sample forms the typical scattering pattern of a polycrystalline solid with bcc structure which is Miller indexed as indicated. Inset: plot of $q^2$ versus $h^2+k^2+l^2$. The solid line is a fit to the data yielding a number density of $n = 1.9$ µm$^{-3}$. c) Thoroughly deionized sample one hour after last homogenization using a rotating tumbler. Again, the scattering pattern can be Miller indexed as indicated to a polycrystalline bcc structure. Inset: plot of $q^2$ versus $h^2+k^2+l^2$. The solid line is a fit to the data yielding a number density of $n = 4.8$ µm$^{-3}$. We note, that due to multiple scattering, a strongly increased background intensity is present at $q > 20$ µm$^{-1}$ in all measurements at large $n$.

The evolution of the static light scattering (SLS) pattern is shown in Fig. 1 for two samples. Fig. 1a shows a thoroughly deionized suspension with number density $n = 1.9$ µm$^{-3}$ at different times after last gentle shearing. A first and second peak can be identified and attributed to the presence of short-range order only. A split in the second peak, as known from silica particle suspensions in the meta-stable melt[37] and also reported for a charged sphere mixture forming an amorphous solid,[42] is clearly absent. Over 24 h no significant structural change occurred; but after two days, we observed slow crystallization *via* homogeneous nucleation. After 3d, the sample had converted to a polycrystalline morphology with bcc structure which can be conveniently Miller indexed (Fig. 1b). A plot of $q^2$ versus $h^2+k^2+l^2$ yields data arranged on a straight line. Its slope yields the number density as $n = 1.9$ µm$^{-3}$. Equivalent scattering patterns could also be obtained in all other samples after sufficiently long waiting times. For comparison, we also show the scattering pattern recorded at $t_w = 1$ h for a sample which was deionized on a rotating tumbler for 3.5 days (Fig. 1c). This sample crystallized instantaneously into a polycrystalline bcc structure *via* heterogeneous nucleation on ion exchange resin debris.

Fig. 1a clearly shows the absence of crystalline long-range order. Similar scattering patterns were observed for the other samples (at shifted $q$ values) before the onset of crystallization. The "lifetime" of such amorphous structures was observed to decrease from days to minutes with increasing particle concentration. No, or at least no significant change of the SLS patterns between the melt state and the amorphous solid state is one hallmark of practically all colloidal and other glasses.[1-5,9] The second indispensable criterion is a finite shear rigidity which can be determined from TRS measurements.

### Elasticity

The existence of a finite shear modulus, $G$ allowed the chance discovery of the amorphous solid in a suspension of PnBAPS118, when some resin splinters did not sediment properly after cessation of shear (see Fig. S2 in Supporting information (SI)).[33] For the samples of this study, *G* is below 1 Pa and shows a systematic increase with *n* in agreement with previous reports from literature.[41,46] At $n = 0.2$ µm$^{-3}$ it amounts to $0.02 \pm 0.006$ Pa[33] and at $n = 0.4$ µm$^{-3}$ it is $0.034 \pm 0.007$ Pa. At all densities the *G* values of the amorphous solid are systematically larger than those of the polycrystalline bcc solids. However at low densities ($n \leq 1.9$ µm$^{-3}$, see Figs. S3 and S4 in SI), no significant discrimination can be made between the shear moduli of samples with different structures. Thus, measurements comparing amorphous to crystalline samples were repeated at enlarged number densities. Results for all densities are shown in Fig. 2. We first fitted the crystalline data with Eqn. (3a) using the number densities from SLS and an estimated background salt concentration of $c = 10^{-2}$ µmol/L (see below in Materials and Methods). This returned an effective elasticity charge of $Z_{eff,G} = 379 \pm 10$. These values then were used in Eqns. (3a) and (3b) to predict the shear modulus of the amorphous solids assuming homogeneously distributed stress with $f_A = 0.6$ and two different types of short-range order. The measured shear moduli of the amorphous samples are consistent with the prediction for the amorphous solid of bcc short-range order. Bcc short-range order is rarely found for colloidal glasses.[46] Most systems, and in particular hard sphere glasses, show a close packed short-range order. Like in hard sphere systems, however, the short-range order of our amorphous solid corresponds to that of the underlying stable crystalline phase.

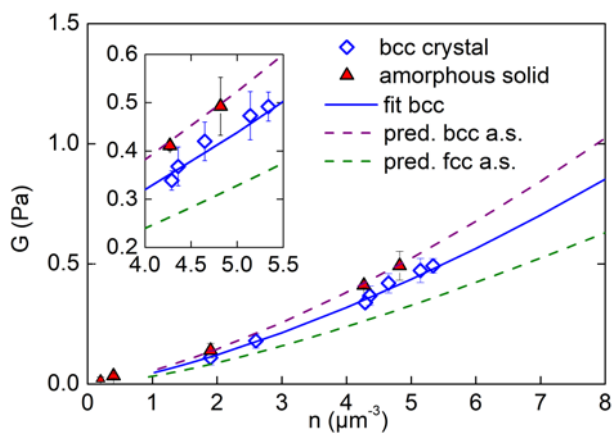

**Figure 2. Elasticity measurements.** Measured shear moduli, *G,* for transient amorphous solids (triangles) and stable bcc crystals (diamonds) as a function of particle number density as compared to the theoretically expected *n*-dependence. (The lowest lying data point was replotted from [33].) The shear rigidity of the amorphous solid is systematically larger than that of the polycrystalline bcc solid. The solid line is a fit of Eqn. (3a) to the crystal data returning an effective charge number of $Z_{eff,G} = 379 \pm 10$. The dashed lines give the predictions by Eqn. (3b) for amorphous solids using this $Z_{eff,G}$ and assuming homogeneously distributed stresses and either bcc (purple) or fcc (olive) short-range order. Inset: enlarged region at elevated *n*, where the difference becomes statistically significant.

**Dynamic light scattering**

Above, we confirmed the existence of an amorphous solid phase in low density suspensions of charged spheres from SLS and TRS data. We now present additional data on the sample dynamics. A characteristic feature of many (colloidal) glasses is the existence of a second slow relaxation process.[6] This has been extensively demonstrated and detailed for hard sphere and charged colloidal systems.[2,19-22,25,29,30,43,44] Also for charged spheres at moderate to low densities, detailed predictions for the evolution of the intermediate scattering function with increasing number density exist.[49,50,53,60] Unfortunately, for the present samples, use of conventional dynamic light scattering (DLS) is inadequate due to multiple scattering effects that yield an ill-defined intercept and plateau value as well as the comparably fast structural evolution which hinders the determination of statistically valid time averages. Note that such effects are practically absent in amorphous suspensions of clay platelets where the intermediate scattering function could be accessed and much information was obtained about the evolution of the sample dynamics.[61-63] Here, however, restricted to the conventional DLS, our means of determining the intermediate scattering function for our steadily evolving non-ergodic samples showing a non-negligible amount of multiple scattering are technically restricted. We nevertheless present intensity autocorrelation functions measured at different *n* in Fig. 3 and Fig.

S5 in the SI. Fig. 3 shows $g^2(q,t)-1$ plotted in a double-logarithmic fashion for a solidifying sample at $n = 1.9$ µm$^{-3}$ ($\Phi = 1.3 \times 10^{-3}$) measured at the primary peak $q_{MAX} = 8.46$ µm$^{-1}$. The initial fast de-correlation due to multiple scattering is integrated out by arbitrarily setting the shortest lag time to 1 µs. Two relaxation processes are clearly visible with a plateau developing in between. The slope of the first one was found to be barely reproducible but the time constant stayed roughly in the range of 0.1 to 1 ms. With increasing waiting time, the second relaxation process appears to slow down and increase in amplitude. A stable functional shape is reached for this sample after about two hours. This "vitrification time" is on the order of a day at $n = 0.4$ µm$^{-3}$ and falls below an hour for $n > 5$ µm$^{-3}$ (Fig. S5 in SI). Over the same range of densities, the onset of crystallization shifts from days to a few hours. Above $n = 8$ µm$^{-3}$, we could not reliably measure the dynamics, since the samples started crystallizing already during the DLS measurements.

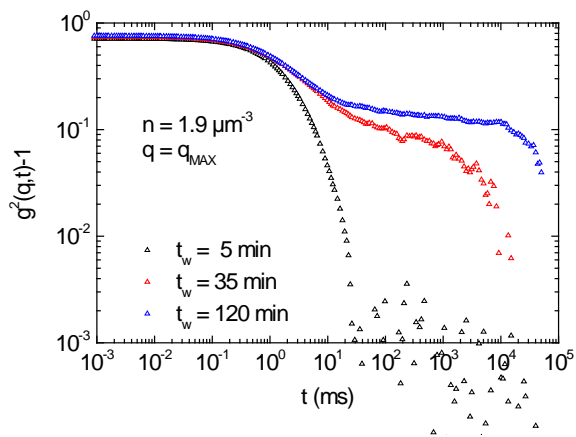

**Figure 3. Normalized intensity autocorrelation function.** Data are plotted in a double-logarithmic fashion for the vitrifying sample at $n = 1.9$ µm$^{-3}$ ($\Phi = 1.3 \times 10^{-3}$) from measurements started after different waiting times as indicated. Note the low intercept stemming from arbitrarily setting the shortest lag time to 0.1 µs, integrating out the fast fluctuations due to multiple scattering. Note further its variability due to non-availability of proper ensemble averaging. Qualitatively, however, two distinct relaxation processes can be reproducibly monitored.

Our findings bear a strong resemblance to the observations made elsewhere on hard[20,21] and charged sphere samples[26] as well as on charged platelets.[25,55-58,61-63,76] There, however, the shape of the intermediate scattering function typically changed in a different fashion. Instead of developing a plateau already at early times which then increases in amplitude, an additional second relaxation process of high amplitude emerged from the first fast one or the slope of the first decay decreased and then a second decay stretched out until a plateau was formed. In fact, for clay platelets, the second relaxation process was well described by a stretched exponential. A highly complex shape evolution of the intermediate scattering function has been observed for multi-arm star polymers.[77] The present shape evolution rather resembles the evolution observable upon nano-crystal formation by homogeneous nucleation,[40,46] or the formation of extended regions of slowed dynamics seen in hard spheres.[22] Given the preliminary nature of our DLS data, we refrain from speculations about any underlying processes. Additional measurements with scattering techniques giving more direct access to the intermediate scattering function or with high resolution microscopy will be needed to decide this interesting point.

**Phase behaviour**

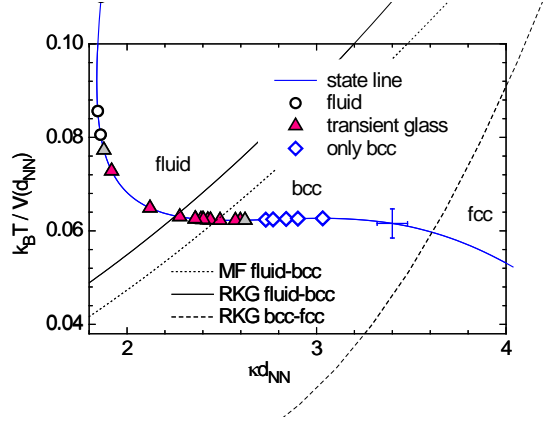

**Figure 4. Experimental phase behaviour in the effective temperature – reduced screening parameter plane.** The thick blue state line shows the path taken by our samples upon increasing the number density (from top left to bottom right) at constant electrolyte concentration and particle effective elasticity charge. The error cross denotes the systematic uncertainties in its location. Thin black lines denote the theoretically expected melting transition and solid-solid phase transition from bcc to fcc taken from different theoretical studies (RKG: ref. [78]; MF: ref. [79] Symbols denote locations of experimental samples. Dark red triangles indicate glasses identified via their liquid-like structure factor and a finite shear modulus, light grey triangles denote samples, with less certain instrumental identification.

Fig. 4. summarizes the results and shows the phase behaviour of deionized PnBAPS118 suspensions as obtained for number densities, $n$, increasing from left to right. Data are shown in the effective temperature – reduced screening parameter plane.[78] Here, the effective temperature, $T_{eff} = k_B T/V(d_{NN})$, is the ratio of thermal energy, $k_B T$, and a hard core Yukawa pair interaction energy:

$$V(r) = Z_{eff,G}^2 \lambda_B k_B T \left( \frac{\exp(\kappa a)}{1+\kappa a} \right)^2 \frac{\exp(-\kappa r)}{r} \qquad (1)$$

where the Bjerrum length is $\lambda_B = e^2/4\pi\varepsilon\varepsilon_0 k_B T = 0.72$ nm in deionized water. Further, $\varepsilon_0$ is the dielectric permittivity of vacuum, $\varepsilon$ is the dielectric constant of the solvent, $e$ is the elementary charge, and the particle hard core radius is denoted as $a$. The distance $r$ is set to the average nearest neighbour distance, $d_{NN}$, estimated from static light scattering as $d_{NN} \approx n^{-1/3}$. The screening parameter is given by

$$\kappa = \sqrt{4\pi\lambda_B k_B T \left( n Z_{eff,G} + n_{salt} \right)} \qquad (2)$$

where $nZ_{eff,G}$ accounts for the effective counter-ion concentration and $n_{salt}$ for the residual electrolyte concentration, estimated to be $c = 10^{-2}$ µmol/L as well as the self dissociation of water of approximately 0.2 µmol/L. The reduced distance, k, is given as the ratio of the nearest neighbour distance and the Debye screening length, $\kappa^{-1}$, as $k = \kappa d_{NN}$. The state line gives possible sample locations for varying number density, $n$, in the thoroughly deionized state as calculated for a constant effective charge number, $Z_{eff,G} = 379$. The uncertainty in positioning the state line is shown by the error bar. The theoretically expected melting lines of Refs. 78 and 79 as well as the bcc-fcc transition of Ref. 78 are shown by lines as indicated. With $n$ increasing from top left to bottom right, the state line crosses these phase boundaries. Actual sample positions are indicated by symbols.

At low densities, the system is in a stable fluid state. The development of a second relaxation process in dynamic light scattering is occasionally seen after one day at $n = 0.15$ µm$^{-3}$ ($\Phi = 1.4 \times 10^{-4}$) and becomes regularly observable at $n = 0.4$ µm$^{-3}$. For $0.2$ µm$^{-3} \leq n \leq 12$ µm$^{-3}$ ($1.9 \times 10^{-4} \leq \Phi \leq 0.01$) the amorphous state was identified by visual inspection through the absence of discrete Bragg reflections and the trapping of dust or minute ion exchange resin splinters in the bulk of the sample. An unequivocal instrumental verification by combined measurements of the shear modulus and the static structure factor was possible in the range of $0.2$ µm$^{-3} \leq n \leq 8$ µm$^{-3}$ ($1.9 \times 10^{-4} \leq \Phi \leq 7.8 \times 10^{-3}$). This range is shown in Fig. 4 by red triangles. Below, samples were too fragile and readily shear melted again; above, samples crystallized too quickly to be unambiguously identified as amorphous solids. Samples which were judged to form an amorphous solid only from visual inspection or inference from the DLS signal are denoted by grey triangles.

For all samples of $n \geq 0.2$ µm$^{-3}$, the amorphous solid was meta-stable against crystallization into a polycrystalline state of bcc structure. For our sample of PI = 0.05 we actually expect an extended fluid-bcc coexistence range.[79] This could, however, not be determined unambiguously, since samples with 0.008 µm$^{-3}$ < $n$ < 0.2 µm$^{-3}$ crystallized only after vigorous shaking or use of the rotating tumbler. Both introduce heterogeneous nucleation at resin splinters mediated by a local increase of the number density at the splinters due to diffusio-phoretic effects.[80] The exact locations of freezing and melting therefore remain to be determined for the fluid-bcc phase boundary.

**Discussion**

A first striking observation from Fig. 4 is the location of the transient amorphous state at or at least very close to the melting transition. The second is its restriction towards high densities by a drastic increase of the homogeneous crystal nucleation rate density. We first discuss the location. Also for charged platelets the glass and gel phases directly border to the ergodic fluid phase. However, crystallization is absent.[25,55-58] The same applies to binary mixtures of hard or charged spherical particles forming eutectics. These show a glass transition within the eutectic gap in the absence of competition with crystal formation,[43,81] similar to the famous Kob-Anderson Lennard-Jones sphere mixtures.[82,83] By contrast, for mixtures forming substitutional alloys,[41,45,84] as well as for one-component systems of both hard and charged spheres,[18,26,31,44] the glass transition is located far above the melting transition. It is typically found at elevated volume fractions with an extended crystalline phase between the freezing transition and the glass transition.[33] Our transient amorphous solids are therefore remarkably different from the colloidal glasses observed in previous experimental studies.

On the theoretical side, glass transition lines are available from several investigations based on mode coupling theory (MCT)[10] using different descriptions for the electrostatic repulsion. In a recent study, Yazdi and co-workers investigated the glass transition in comparison to the melting transition for charged colloids and complex plasmas. These authors used a Yukawa description of point charges for the pair interaction and their results are conveniently plotted in the coupling parameter – reduced distance plane. In the limit of vanishing screening ($\kappa \to 0$), the melting transition of the one component plasma (OCP) is located at a coupling parameter $\Gamma_{F,OCP} = V(r)/k_B T \exp(\kappa r) \approx 106$,[78,79,85] where $V(r)$ is the pair interaction energy and $r$ is set to $r = d_{NN}$. The MCT glass transition is located at $\Gamma_{MCT,OCP} = 366$. When screening is introduced for $\kappa > 0$, both the melting and the glass transition move to higher critical coupling strengths as $\Gamma(\kappa) = \Gamma_{OCP} \, e^{\kappa}/(1 + \kappa + \kappa^2/2)$, *i.e.*, the glass line is predicted to run parallel to the freezing line in a plot of $\Gamma$ versus k.

We compare our data to their calculations in Fig. 5. With $n$ increasing from bottom left to top right, the state line – calculated using Eqns. (1) and (2) with the same parameters as in Fig. 4 – crosses first the melting line then the glass line. The amorphous solid is observed very close to melting but somewhat on the side of the fluid phase. All samples marked by red triangles actually convert to fully crystalline samples. Amorphous solids become too short-lived to be observed as the density is increased significantly above the melting density. No reliable observations could be made at or above the predicted glass line due to turbidity issues. The prediction for melting is met reasonably well, given the experimental uncertainties involved in the localization of the state line. Moreover, also previous work on charged sphere samples of moderate polydispersity under thoroughly deionized conditions,[69,70] occasionally showed a somewhat larger than predicted stability of experimental crystals. The agreement of the locations of experimental amorphous solids and the range of parameters for which the point-Yukawa MCT calculations predict a glass is less convincing.

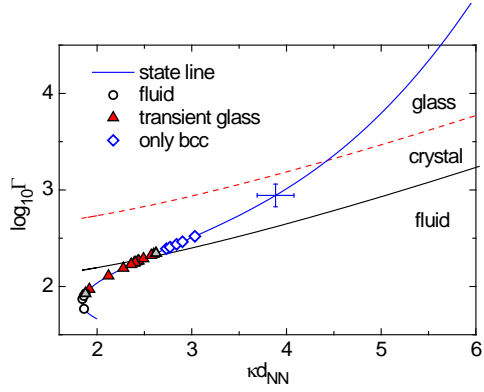

**Figure 5.** Comparison of experimental phase behaviour to the predictions of Yazdi et al.[60] Melting line (solid) and MCT glass transition line (dashed) are shown in a semilog plot of the coupling parameter – reduced distance plane. The thick blue line shows the path taken by our samples (state line) upon increasing the number density (from bottom left to top right) at constant electrolyte concentration and particle effective elasticity charge. The error cross denotes the systematic uncertainties in its localization. Symbols as before in Fig. 4. The theoretical glass and melting lines underestimate the stabilities of amorphous and crystalline solids. The experimental amorphous solids are observed directly adjacent to the fluid phase.

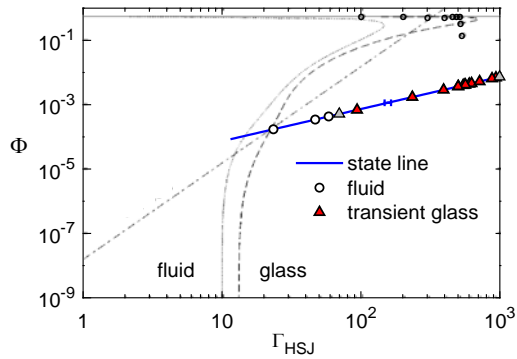

**Figure 6.** Comparison of experimental phase behaviour to the predictions of Wilke and Bosse.[52] Data are shown in a double logarithmic fashion in the volume fraction – plasma parameter plane. The thick blue line shows the path taken by our samples (state line) upon increasing the number density (from bottom left to top right) at constant electrolyte concentration and particle effective elasticity charge. The error cross denotes the systematic uncertainties in its localization. Symbols as before in Fig. 4. Thin black lines denote the theoretical results for the location of the MCT glass transition. Solid line: charge neutral hard spheres; dotted line: restricted primitive model (RPM); dashed line: hard sphere jellium model (HSJ). The dash dotted line represents a line of constant $T^* = 0.005$. Small circles are glass transition points reproduced from Lai and Chang.[86] Experimental data are well compatible with these predictions, however, the predictions overestimate the stability of the amorphous solid. (Theoretical data reprinted with permission from [52]. Copyright (1999) by the American Physical Society.)

Next, we compare to the MCT predictions made for charge neutral hard spheres, the hard sphere jellium model (HSJ) and the restricted primitive model (RPM) as reported by Wilke and Bosse,[52] including some data on charged hard spheres of Lai and Chang.[86] Since all these interaction types comprise a hard core, the authors plot their results in the volume fraction – plasma parameter plane. The plasma parameter is defined as $\Gamma_{HSJ} = 2\Phi^{1/3}/T^*$, with a reduced temperature $T^* = k_B T 8\pi\varepsilon_0 a/(Z_G e)^2$. HSJ and RPM predict a re-entrant glass behaviour. Charged sphere glasses are predicted to be stable also at very low volume fractions, given $\Gamma_{HSJ} \geq 10$. To plot our data, we identify the particle radius with $a_{SAXS}$ and the charge with $Z_{eff,G}$ and calculate $\Phi$ using $a_{SAXS}$ and the experimentally determined $n$. Our experimental data are consistent with the predictions, which, however, seem to overestimate its stability.

The very existence of an amorphous solid at low number density of charged spheres qualitatively confirms the predictions of MCT based calculations. The theoretical approaches differ in their choice of interaction potentials and their results. A discrimination among these is not yet feasible with the present data set based on the use of $Z_{eff,G}$ and a single value for the salt concentration. $Z_{eff,G}$ accounts correctly for many body effects in crystalline solids.[69] Its applicability for the description of our amorphous solid is supported by the agreement between predicted and measured shear moduli. A continuous adjustment of the salt concentration could be obtained using other preparation techniques.[87] Finally, a better localization of the freezing and melting transitions appears to be necessary, which may be achieved by crystal growth experiments using heterogeneous nucleation.[79]

However, MCT does *not* predict an upper density limit for vitrification. As in the case of buoyancy matched hard sphere glasses, it fails in explaining the retained particle mobility which allows crystallization from the amorphous state.[8,21,31] In fact, for the present amorphous solids, the observation limit at large meta-stability is given by the rapid crystallization. For all samples, the lifetime of the amorphous colloidal solid decreases from days to minutes with increasing particle concentration. It is given by the span between its formation from the shear melt and the onset of crystallization. During formation, a certain time span is required to reach stationary values for the shear rigidity and the relaxation time for the slow process. This time span decreases with increasing $n$. Inspection of our preliminary DLS measurements seems to suggest that the final value for the relaxation time of the slow process is smaller at larger density. Further, crystals form ever faster and appear earlier as $n$ is increased. Hence, both the way into and out of the amorphous state speed up with increasing particle concentration but the acceleration of transformation kinetics appears to become more pronounced for crystallization. This leads to an ever shorter lifespan of the amorphous state.

Our observations are therefore qualitatively consistent with expectations based on a strong electrostatic coupling between the particles and the standard nucleation theory. A few points, however, need additional attention. Possibly, the most striking feature of our observations is its uniqueness. Despite numerous solidification experiments on very similar systems, this is the first charged sphere system to show such a low density amorphous solid. In fact, it was only discovered by chance in samples crystallizing very reluctantly. If we attribute the observed formation of an amorphous solid to the presence of strong and weakly screened electrostatic interactions, we have to wonder why in previous studies, deionized aqueous suspensions of highly charged spheres with sizes between 50 nm and 300 nm and polydispersity indices in the range of $0.02 \leq PI \leq 0.11$ all crystallized.[65-70] Moreover, their freezing points in the effective temperature – coupling parameter plane were observed to be close to each other and in the same parameter range as in the present case. Thus, while we believe that a strong mutual interaction is a necessary condition to enable the formation of an amorphous state, it seems not to be a sufficient one.

A possibly sufficient criterion may be a simultaneous and effective suppression of nucleation. This is observed in systems of clay platelets due to their non-spherical interactions disfavouring simple packings. However, spheres with only slight shape and interaction asymmetry have been found to crystallize at low densities in absence of any competition with glass formation.[88] Therefore, we expect the near perfect spherical symmetry of our particles apparent from TEM and SAXS (see Fig. S1 in SI) to render the interactions spherically symmetric. We therefore exclude this as the reason for a possible slowing of nucleation. According to the classical nucleation theory, also a large nucleus-melt interfacial energy can suppress crystal nucleation. In principle, polydispersity could make this difference. In a recent study we observed a clear anti-correlation between size-polydispersity and the interfacial free energy involved in homogeneous nucleation.[89] Therefore, we carefully checked the low nominal polydispersity index of the present system of $PI = 0.011$ with several additional measurements to find a corrected value of $PI \approx 0.05$ (Fig. S1). This excludes an exceptionally low polydispersity to be responsible for a suppression of nucleation and the formation of amorphous solids. Our additional experiments also exclude an exceptionally large polydispersity. These may lead to fractionation effects, which are known to increase the melt-nucleus interfacial free energy,[90,91] require composition fluctuations and therefore drastically decrease nucleation rate densities.[81,92] The reason for the apparently odd behaviour of PnBAPS118 therefore remains completely obscured, as this species as well as the present preparation methods are unsuspicious in any way.

A second interesting point concerns the way into the amorphous state and the involved mechanisms. Like in a high density system of charged spheres reported by Schöpe et al.,[46] the present amorphous solid shows a bcc short-range order as evidenced by the shear modulus measurements on solidified samples. The solidification time scale, however, is markedly different. At high density, the solidification process could not be resolved in time. The present samples solidify on well measurable time scales. Slow solidification (from the homogenized shear melt) is also known from hard sphere and charged platelet glasses[25,38,76] and the associated structural development has there been termed aging. On the theoretical side, aging has been associated with an exploration of configuration space in which the samples reach ever deeper local meta-stable minima in free energy, while experiencing a progressively slowing of their (activated) dynamics. Irrespective of the underlying mechanism (individual hopping events or collective rearrangements) the ageing process results in a gradual decrease of overall particle mobility and an increase in resistance to mechanical stress. In dynamic light scattering it shows up as a second slow relaxation process as also observed here.

It has long been recognized in glasses and gels, that dynamical slowing down typically is associated to spatially localized, dynamical heterogeneities as well as sub-diffusive dynamics.[93-95] Other studies suggest the cease of purely diffusive motion as soon as the meta-stable melt is reached[96] and the arrest of flow combined with the onset of hopping processes at the glass transition density.[97] The exact nature of the underlying microscopic particle motion and its connection to structural heterogeneity, as well as to the structure of the energy landscape are still under discussion[4,6-8,11,15,17,22,30,98] We explicitly note that our observation of a second slow relaxation process does not exclude the existence of spatially heterogeneous dynamics but can, of course, not confirm these. For that, additional measurements employing e.g. confocal microscopy,[99] differential dynamic microscopy[100] or spatially resolved multi-speckle correlation spectroscopy[22] are necessary.

An alternative explanation for the emergence of a second slow relaxation process in our meta-stable melts could be given by the formation and intersection of locally preferred structures (LPS) or medium-range crystalline ordered (MRCO) structures linking to structural heterogeneity associated with differing particle mobilities.[101] LPS break symmetry and at the same time are mutually incompatible such that they are unable to tile space completely. Examples include Bernal glasses,[16] icosahedral structures or clusters of related symmetry.[34,35,37] Frustration may further result from orientational incompatibility of MRCO regions or sub-critical nuclei.[11,46] Also such strongly ordered precursor structures could result in a second relaxation process, e.g. from their own much slower diffusion as a whole.[102] Formation of amorphous solids then further requires the intersection and jamming of these structures[103] which otherwise would stay mobile and leave the melt in an overall liquid state.[37] Jamming impedes their growth and additionally slows the melt dynamics in the intersection regions.[7,8,11] Note that also such a state would be compatible with the observation of two distinct relaxation times. A stabilization of the amorphous state by intersecting locally well ordered regions obtains experimental support by several studies on hard or charged spheres.[29,34,35,46] Formation of MRCOs, however, still allows for subsequent crystallization. In this respect it is interesting to note that in many systems crystal nucleation proceeds via a two-step scenario[104-106] involving the formation of crystal precursors and of two different order parameters: symmetry and density.[107] For hard-sphere systems, experimental studies of dynamic heterogeneities in supercooled melts show that there is a close match between slow regions and regions of increased density[22] or with high crystal-like bond-orientational order.[34] Studies using confocal microscopy have in addition revealed that both conditions also apply for fcc crystal precursors in slightly charged hard sphere melts.[105] For low density systems crystallizing bcc, it was observed that the crystals can form *via* bcc pre-ordered low density precursors.[108] In the present charged sphere systems, the transient amorphous solids possess bcc short-range order, which is retained in the nucleating crystals. Therefore, the present observations of slowing dynamics at unchanged overall structure and density seem well compatible with the formation of MRCO regions or precursors of bcc-like structure by an activated process followed by their intersection resulting in the formation of a transient amorphous solid. High resolution microscopy employing fluorescently labelled particles could be used to test this suggestion directly in real space.

Assuming bcc-LPS, MRCO or pre-cursor formation to be a general process in colloidal solidification again raises the question, why it has always lead to crystals in previous studies, but here should be involved in the formation of a transient amorphous solid. Here, one may only speculate. Possibly, the particular combination of effective charge, particle size and distance as well as the very low background salinity facilitates a particularly pronounced short range order in the melt state. Let's in addition assume that such regions are smaller than the sizes of critical nuclei for which classical nucleation theory predicts a significant decrease in size with increased meta-stability. Given the strong electrostatic interaction and the slow diffusion dynamics of the melt, differently oriented bcc-like regions would be very slow to reorient and coalesce. Only coalescence, however, would keep the structure while enhancing the extension of regions of high bond orientational order. Recurring coalescence eventually may yield a post-critical nucleus ready to grow. Systematic measurements to study this system at other electrolyte concentrations and of similarly sized but differently charged particles are under preparation. Ideally, parameters and experimental boundary conditions allowing the formation of amorphous solids at low density and of crystals at large density can be identified which apply not only for PnBAPS118 but can be realized in other charged sphere systems, too.

The above discussion indicated two possibilities for glass formation in the present system without any preference. Other approaches may as well turn out to be suitable. Crystal formation from the amorphous solid state leading to a restricted observation range may provide first qualitative constraints as this possibly can be traced back to competing time scales for the formation of amorphous and crystalline solids. Details of the crystallization mechanisms and kinetics may be accessed by time resolved static light scattering or Bragg-microscopy. Nucleation and growth of crystals from the melt state has been intensively studied[79,106,109] and typically is parameterized within classical nucleation theory (CNT)[1,2,110] or related alternative approaches.[111,112] Nucleation from an amorphous solid has so far escaped detailed investigation due to the inconvenient time scales involved. Here, nucleation is outpaced by the formation of the amorphous state at low density, but catches up with increasing density and overtakes at large densities. At intermediate densities, our system offers a convenient possibility to study crystallization from the amorphous state.

## Conclusion

We have clearly demonstrated the existence of an amorphous solid in a model system of low-density charged spheres suspended in thoroughly deionized water. We gave a first systematic characterization combining different optical experiments. The low-density soft glass studied here exhibits significant and interesting differences from the known high-density hard or charged sphere glasses. Its unique features include the location of the amorphous phase with respect to the melting line of the underlying stable bcc phase, the bcc short-range order and the peculiar density dependence for the outcome of the competition between crystallization and vitrification. We discussed possible ways into and out of this novel type of transient amorphous solid. The observations presented raise interesting questions: What are the kinetic pathways into and out of the amorphous solid? How can an amorphous solid form at such low densities? Why is it so strongly affected by the competition with crystallization? And why is this state so elusive? A final discrimination between different theoretical approaches describing the formation of "the" amorphous state in general and the present one in particular requires more detailed observations with complementary experimental methods. We therefore believe that our observations qualify as points of departure for a larger number of investigations on charged sphere amorphous solids. We therefore anticipate that they will stimulate enhanced experimental and theoretical interest on this particular type of amorphous solids and in the long run facilitate systematic tests of theoretical concepts of colloidal glasses.

## Materials and Methods

**Samples and sample preparation**

**Particle characterization:** Particles of Lab code PnBAPS118 are co-polymer latices and were synthesized by emulsion polymerization. They contain a mixture of Poly-n-Butylacrylamide (PnBA) and Polystyrene (PS) at a composition of 35:65 W/W and are stabilized by sulphate surface groups. They were a kind gift of BASF, Ludwigshafen (manufacturer Batch No. 1234/2762/6379). Their nominal hydrodynamic diameter (from routine dynamic light scattering by the manufacturer) was given as $2a_h = (117.6 \pm 0.65)$ nm corresponding to a nominal polydispersity index (standard deviation divided by the mean radius) of PI = 0.011. We re-checked these data by transmission electron microscope (TEM) to obtain PI = 0.047 for cryo-TEM and PI = 0.056 for Negative Staining TEM as well as by small angle X-ray scattering (SAXS) to obtain PI = 0.051 for spheres of average hard core radius $2a_{SAXS} = 109.1$ nm covered by a Gaussian chain layer of thickness $d = 1$ nm. These results are shown in Figs. S1a and S1b.

The effective charge numbers were determined by two different methods.[114] The number density dependent conductivity measurements on deionized samples yield an effective conductivity charge number $Z_{eff,\sigma} = 647 \pm 18$.[113] $Z_{eff,\sigma}$ corresponds to the number of freely moving counter ions, coincides well with Poisson–Boltzmann cell model calculations and accounts for self-screening, charge regulation and charge renormalization.[75] Elasticity measurements using Torsional Resonance Spectroscopy (TRS) interpreted in terms of an effective charged hard sphere pair potential yield an effective elasticity charge $Z_{eff,G} = 379 \pm 10$.[114] $Z_{eff,G}$ in addition accounts for many body effects known as macro-ion shielding.[115] It therefore is systematically lower than $Z_{eff,\sigma}$ and yields an excellent consistency of experimental and expected melting line locations.[69,70]

**Sample conditioning:** Since the vitrification experiments have a long duration, it is crucial to precisely maintain the electrolyte concentration over several days. This excludes conditioning in so-called continuous deionization which removes dissolved ions, but is not gas tight over extended times after stopping the cycling of the sample through mixed bed ion exchange resin. Therefore, in continuous deionization, the electrolyte level slowly increases again due to the dissolution of gaseous $CO_2$ and dissociation of the carbonic acid.[87] We therefore adapted Okubo's method of batch preparation.[45,67] The supplied stock suspension ($\Phi \approx 0.2$, $n \approx 230$ μm$^{-3} = 2.3 \times 10^{20}$ m$^{-3}$) was first diluted and stored over mixed bed ion exchange resin (Amberlite, Rohm & Haas, France) for a few weeks under occasional gentle stirring. It was then filtered to remove dust, resin debris and coagulate, regularly occurring upon first contact with the exchange resin. The procedure was repeated twice using fresh resins and the cleaned stock solution was then stored in a fridge. From this stock, samples of desired number densities (corresponding to volume fractions of 0.02 and less) were prepared by dilution in 2 mL sample vials with freshly rinsed ion exchange resin added. Samples were sealed against airborne $CO_2$ with Teflon® septum screw caps (Sigma Aldrich, Germany). They were left for more than two months until the crystallite sizes obtained after shear melting became constant, indicating thoroughly deionized conditions.[45] Great care was taken to use only very gentle turning and slow shaking for melting to avoid the formation of resin debris which acts as heterogeneous nucleus for crystallization. Samples shaken too hard were identified by fast crystallization (c. f. Fig. 1c). These had to be filtered and deionized again before reuse. Experiments were performed in dependence on number density and waiting time $t_w$ defined as the time after last gentle shaking. First scattering measurements typically started with a delay of 3 min needed for sample mounting.

**Cryo-TEM:** Specimens were prepared in a modified cryo plunge type CP3 (Gatan, USA) to ensure fixed temperature and avoid water loss from the sample during preparation. In the cryo plunge, 5 μL droplet of stock suspension was placed on a carbon S160 coated TEM-grid (Plano GmbH, Germany), which was surface activated by oxygen plasma for 30 s before use. It was then soaked by a filter paper for 1.5 s at a humidity of higher than 95 %, resulting in the formation of a thin liquid film. Afterwards the grid was instantaneously shot into liquid ethane at its freezing point leading to vitrified specimens. Subsequently, the vitrified specimens were transferred to the Tecnai 12 TEM (FEI, USA) using a Gatan 626 cryoholder and its "workstation". Imaging was carried out at a temperature of about −170 °C and 100 kV acceleration voltage.

**Negative Staining TEM:** Specimens were prepared from a 5 μL drop of dilute sample solution adsorbed to an oxygen plasma activated carbon-coated copper grid and stained with 5 μL of 2 % uranyl acetate solution for several minutes.

Subsequently, the liquid was adsorbed by filter paper. Specimens were imaged at room temperature using a Tecnai G2 Spirit TEM (FEI, USA) at an acceleration voltage of 120 kV. An example is shown in the inset of Fig. S1a.

**SAXS:** The Xeuss SAXS/WAXS system (Xenocs, Sassenage, France) features a 30 W Cu Kα microfocus tube with ultra low divergence mirror optics (GeniX, Sassenage, France) and a Pilatus 300K/20Hz hybrid pixel detector (DECTRIS, Baden Dättwil, Switzerland). Focus area at the sample is 0.6 mm² for high resolution collimation. The motorized components were controlled with SPEC software. The samples were filled into a Kapton flow-through capillary (inner diameter 1 mm, wall thickness ± 0.025 mm (Goodfellow GmbH, Germany)) mounted on a Linkam stage (Linkam Scientific Instruments, United Kingdom) kept at room temperature. The sample-to-detector distance for SAXS was 2770 mm, calibrated with silver behenate. The absolute calibration of the scattering data was done with glassy carbon type 2, sample P11.[116] The X-ray scattering vector $q$ is defined as $q = 4\pi/\lambda \sin(\theta)$ at a scattering angle of $2\theta$. Data were fitted by several model functions for smooth hard spheres, fuzzy spheres and hard spheres decorated with a layer of Gaussian chains.[117] Only the latter fit function was able to describe the data over the full range of scattering vectors down to the noise level.

**Optical experiments**

A first characterization of our samples was always performed by visual inspection. Measurements to determine the sample structure, elasticity and dynamics were all performed using a multi-purpose light scattering instrument described in detail elsewhere.[40] This instrument allows quasi simultaneous measurements of sample structure by static light scattering (SLS) and sample dynamics by dynamic light scattering (DLS) as well as of elasticity by TRS without the need to transfer the fragile samples from set-up to set-up.

**Visual inspection:** A first check of sample structure can be performed by simple visual inspection. Liquid-like structure is identified from the absence of Bragg reflections, while crystalline structure from their presence. Observation of non-settling dust particles and in particular non-sedimenting ion exchange resin debris evidences a finite shear rigidity. Photographs of representative samples taken at different number densities are reproduced in Fig. S2.

**SLS:** Laser light of wavelength λ = 647.1 nm is alternatively fed in two optical fibres and sent counter-propagating through optics optimized for SLS and DLS into the sample. Scattered light is picked up by receiving optics mounted on two opposing goniometer arms, hence recording SLS under the same scattering vector. To capture possible fast changes in structure, SLS was recorded in fast mode with the stepper motor covering 157° in 240 steps with 3 s integration time each. We further refrained from calculating the static structure factor because of the unknown amount of q-, structure- and density dependent multiply scattered light. This forbids the conventional calculation of $S(q)_{ordered} = I(q)_{ordered} / I(q)_{disordered}$ by division of $I(q)_{ordered} = I_0\, n\, b_0^2\, P(q)\, S(q)$ with the intensity pattern of a disordered sample at the same number density $I(q)_{disordered} = I_0\, n\, b_0^2\, P(q)\, 1$. Here $I_0$ is the instrumentally determined detection efficiency, $b_0^2$ is the single particle scattering cross section and $P(q)$ is the form factor. However, also from $I(q)$ we can unequivocally discriminate among fluid-like order and the onset of crystallization in all investigated samples, and, moreover we can accurately determine the sample number density from the late stage crystalline samples (c.f. Figs. 1b and 1c).

**TRS:** Details of the elasticity measurements have been given previously for the case of polycrystalline or single crystalline samples in different geometries.[118-120] For TRS, the static-side optics are used for illumination. The sample cell is set into low-amplitude oscillations about its vertical axis, which excites the eigenfrequencies of the solid in the known cylindrical geometry. A reference signal is obtained from the reflection of a second laser beam off a small mirror fixed to the sample outside and recorded by a position sensitive detector (PSD, SSO-DL100-7, Silicon Sensor, Berlin, Germany). Scattered light is recorded by a second suitably positioned PSD. For crystalline samples, an individual Bragg reflection is chosen and its peak position $(x_{MAX}, y_{MAX})(t)$ is detected as a function of time. For amorphous samples, a scattering vector $q < q_{MAX}$ on the low-$q$ slope of the primary peak in $I(q)$ is selected. The PSD runs in integral mode and the periodic change in the scattered light intensity $I(q,t)$ is monitored. Using a dual channel lock-in amplifier (SR530, SRS, Sunnyvale, CA) the resonance spectrum is recorded for frequency intervals of (0.5–10) Hz. Typical spectra are shown for an amorphous sample at $n = 1.9$ µm$^{-3}$ in Fig. S3a and for the subsequently formed crystalline phase in Fig. S3b. Note the shift of resonance frequencies to lower values after crystallization. Note further, that the modes are more clearly resolved in the crystalline sample.

The positions of the eigenfrequencies then yield the shear modulus $G$ of the sample in dependence on crystal structure.[121] For bcc crystals, $G$ is given as:

$$G_{bcc} = f_A \frac{4}{9} nV(d_{bcc}) \kappa^2 d_{bcc}^2 \text{ with } d_{bcc} = \frac{\sqrt{3}}{\sqrt[3]{4n}} \qquad (3a)$$

while for fcc it reads:

$$G_{fcc} = f_A \frac{1}{2} nV(d_{fcc}) \left( \kappa^2 d_{fcc}^2 - \kappa d_{fcc} - 1 \right) \text{ with } d_{fcc} = \frac{\sqrt[6]{2}}{\sqrt[3]{n}} \qquad (3b)$$

Here, $f_A$ is a numerical factor which accounts for the different boundary conditions in averaging over randomly oriented crystallites or local environs. Its theoretical limits are $f_A = 0.4$ for homogeneously distributed strains and $f_A = 0.6$ for homogeneously distributed stresses, respectively.[122-125] For polycrystalline samples a value of $f_A = 0.5$ is encountered in most cases.[40,41,131] Using a hard core Yukawa pair interaction $V(d)$ in Eqn. (1) we can solve for $Z_{eff,G}$. This characteristic quantity can be used for comparison with the predicted phase behaviour but furthermore, also for predictions of $G$ for glassy samples assuming homogeneously distributed stresses.

The broad resonance lines of amorphous samples yield an increased systematic uncertainty which is estimated to be on the order of about 10% for the low $n$ samples. There we estimate an upper bound for the experimental uncertainty to be about 20%. For samples with $n > 1$ µm$^{-3}$, the total uncertainty is dominated by statistical uncertainties (obtained from repeated measurements under identical conditions) which are on the order of 10% at all densities. The detection limit of our set-up and cell geometry was found to be on the order of 0.01 Pa. It is set by the small shear moduli themselves. At the lowest number densities, freshly solidified samples typically re-melted when going into torsional resonance. Samples aged for half a day could be measured with less difficulty, but altering of $f_A$[129] by plastic deformation cannot be generally excluded. Therefore, the frequency sweep for low $n$ samples was performed from high to low frequency. Still at low frequency, strong resonant vibration often resulted in a structural change of the sample. Such a "glitch" can be identified by a steep decrease of vibration amplitude and a plateau-like feature in the phase. This is shown exemplarily and marked by red arrows in Fig. S3c for a sample at $n = 0.4$ µm$^{-3}$. The measurement was immediately repeated and the subsequent spectrum reveals a shift of resonances to lower frequencies and an overall clearer resolution (Fig. S3d). To support this interpretation we performed measurements of the static light scattering pattern on samples aged for several hours and immediately after performing a TRS experiment with a glitch occurring. An exemplary result is shown in Fig. S4. For shear modulus determination in low density amorphous samples we therefore only used gently shaken samples aged for a sufficiently long time. In addition, only the high frequency part of the spectra recorded before an eventual glitch was used for evaluation.

**DLS:** DLS was recorded by photomultiplier (H5783P, Hamamatsu) and analyzed by a digital correlator with PM-PD-unit (ALV-7004, ALV, Germany). Our samples are strongly affected by multiple scattering.[126] Therefore, unlike in clay suspensions conventional dynamic light scattering here does not allow to obtain statistically reliable data which can be interpreted in a quantitative way and then compared to theoretical models.[55-58,61-63] To at least obtain qualitative data, we therefore arbitrarily set the shortest lag time to 1 µs, integrating out the fast intensity fluctuations due to multiple scattering and we restricted typical measurement durations to times between 20 and 30 min in order to accommodate several successive runs at short $t_w$. Moreover, no ensemble averaging was performed. Both leave the intercept and the plateau at intermediate times and the final baseline ill defined. We therefore only qualitatively state the presence or absence of the second slow relaxation process and give estimates of the evolution of relaxation times with increasing $t_w$, but we refrain from any further interpretation of the intensity autocorrelation function. Fig. 3 in the main text shows a double log plot of $g^2(q,t)$-1 obtained for different $t_w$ at $n = 1.9$ µm$^{-3}$. These data are shown again in a lin/log fashion in Fig. S5c together with examples of $g^2(q,t)$-

1 obtained at $n = 4.8$ µm$^{-3}$ (Figs. S5a and S5b) and $n = 0.4$ µm$^{-3}$ (Figs. S5d and S5e). The qualitative features of i) an ill defined intercept and plateau height, ii) occurrence of a second relaxation process and iii) its apparently systematic shift with time are seen in all cases. Moreover, there appears a clear trend of the relaxation times of slow process to evolve faster at elevated $n$.

We are fully aware of these present limitations to adequately deal with the experimental challenges of fast structural evolution and multiple scattering arising from the high refractive index contrast of our water-based suspensions. Several cross-correlation schemes have been proposed to access static and dynamic data in turbid samples.[118] An additional challenge for polycrystalline or glassy samples is a correct ensemble average in order to proceed from the measured intensity autocorrelation function to the desired intermediate scattering function.[127-134] In the mentioned clay suspensions,[55-58,61-63] the temporal evolution of the sample dynamics was very slow (over several tens of hours). This gave enough time to perform valid time averages on which the Siegert relation could be applied to infer the intermediate scattering function. In the case of charged sphere glasses, alternative approaches based on synthesis of low refractive index polymer latices[44] or heterodyne scattering equipment[135,136] may be necessary to solve both issues simultaneously. The present study therefore has to be considered preliminary in the sense that we can only present data which are not yet corrected for multiple scattering. We anticipate, however, that the qualitative features seen in these additional data also will be recovered in future systematic studies relying on the single scattering intermediate scattering function and multiple scattering corrected static structure factors.

## Acknowledgements
We are pleased to thank Matthias Sperl, Alexander Ivlev, Hartmut Löwen and Kurt Binder for encouraging discussions on Coulomb glasses and Thomas Hellweg for the discussion of SAXS data, Bastian Barton and Frank Depoix for technical assistance in TEM measurements and BASF, Ludwigshafen for the kind gift of these particles. We gratefully acknowledge financial support by the Deutsche Forschungsgemeinschaft (DFG) (Pa459/16, Pa459/17, INST 215/432-1 FUGG) and the JGU (interne FoFö 1. Stufe). S. H. is a recipient of a fellowship through the Excellence Initiative (DFG/GSC 266). R. I. D. is an Erasmus+ Trainee and recipient of a fellowship through the Duitsland Intstituut Amsterdam (DIA).

## Author Contributions
All authors contributed to this work. R. N., R. I. D., & M. H. performed the light scattering measurements, S. H. performed the TEM measurements, R. S. conducted the SAXS measurements and evaluation. R. N. & T. P. wrote the manuscript. All authors reviewed the manuscript.

## Additional information
**Competing financial interests:** The authors declare no competing financial interests

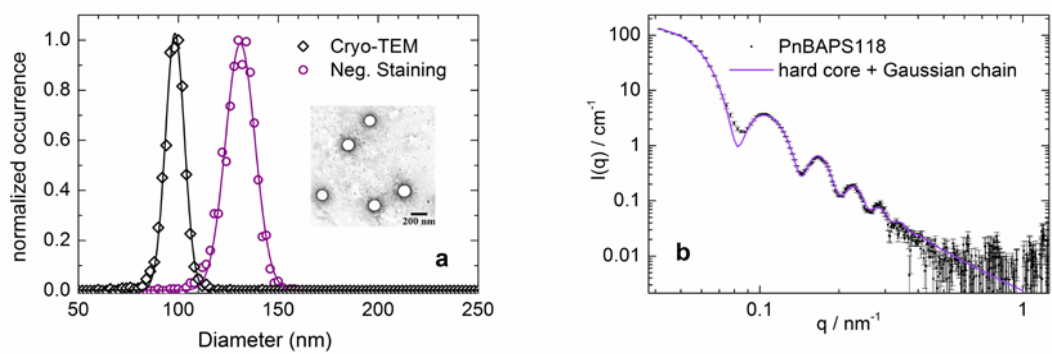

**Figure S1. Size and polydispersity determination.** a) Results of TEM measurements. For each method, some 1500 particles were imaged, tagged manually to exclude doublets and other artefacts. Sizing was done by standard software. Due to a failure of the absolute calibration, both TEM data sets could only be evaluated for relative widths of the distributions. The solid lines are fittings with Gaussian distribution function and the fitted PI are 0.047 and 0.056, for Cryo-TEM and Negative Staining TEM, respectively. The inset shows an example of a Negative Staining TEM image. b) SAXS scattering curve. The solid line is the fit of the hard core plus Gaussian chains model which returns PI = 0.051, $2a_{SAXS}$ = 109.1 nm and $d$ = 1 nm.

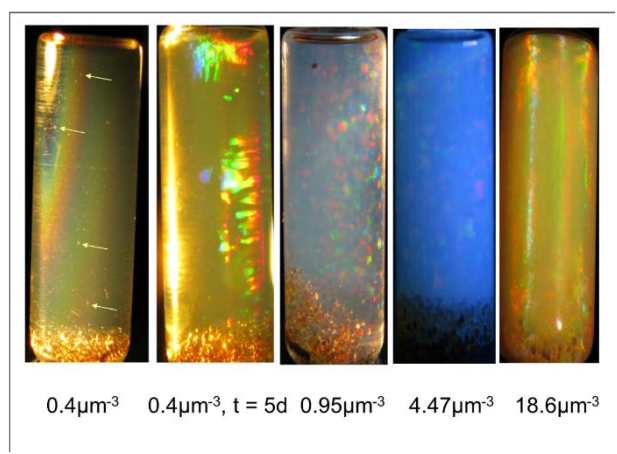

**Figure S2. Images of PnBAPS118 samples** (illuminated from aside). These were taken at different indicated number densities, $n$, increasing from left to right. Sample height is ca. 35 mm. The leftmost sample is an amorphous solid, as recognized from the pronounced double rainbow and the non-sedimenting ion exchange resin splinters marked by the arrows. After five days, columnar crystals have grown into the amorphous solid after heterogeneous nucleation at the cell wall. At larger number densities homogeneously nucleated crystals are obtained. Note the strong multiple scattering present in the samples at larger $n$ ($n > 1$ µm$^{-3}$) which renders these milky to opaque. (reprinted with permission from Ref. S1. © SISSA Medialab Srl. Reproduced by permission of IOP Publishing. All rights reserved.)

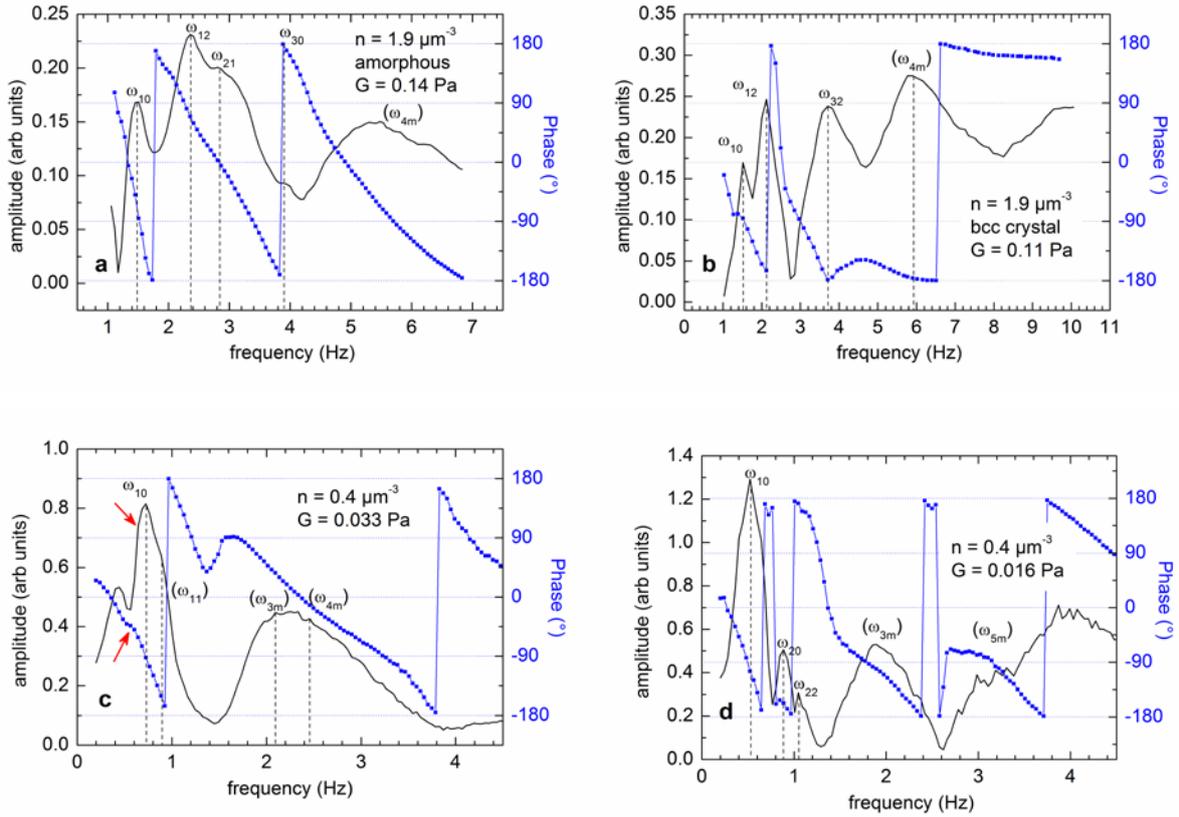

**Figure S3. TRS spectra.** Data shown for samples at $n = 1.9$ µm$^{-3}$ (a and b) and $n = 0.4$ µm$^{-3}$ (c and d) in amorphous (a and c) and crystalline states (b and d). The red arrows in (c) labels a glitch occurring at strong resonant vibration. Immediately afterwards, another frequency sweep was performed and the corresponding spectrum is shown in (d) where the resonance peaks are sharpened but shifted to lower frequencies as compared with (c). The resonance frequencies are identified with the mode indices indicated. The derived shear moduli are indicated in the figures.

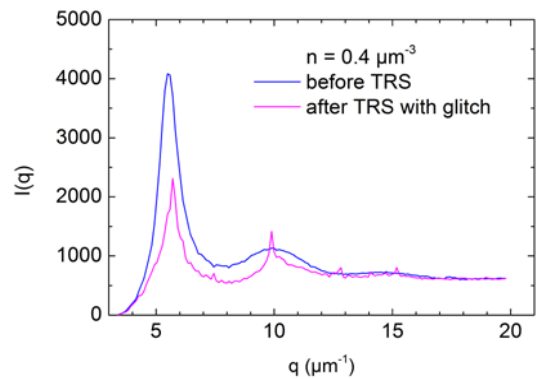

**Figure S4. Static light scattering pattern of a sample at $n = 0.4$ µm$^{-3}$.** Shown are data recorded after leaving the sample undisturbed for several hours and after performing a TRS experiment with a glitch occurring at resonance. The latter data clearly show the onset of crystallization.

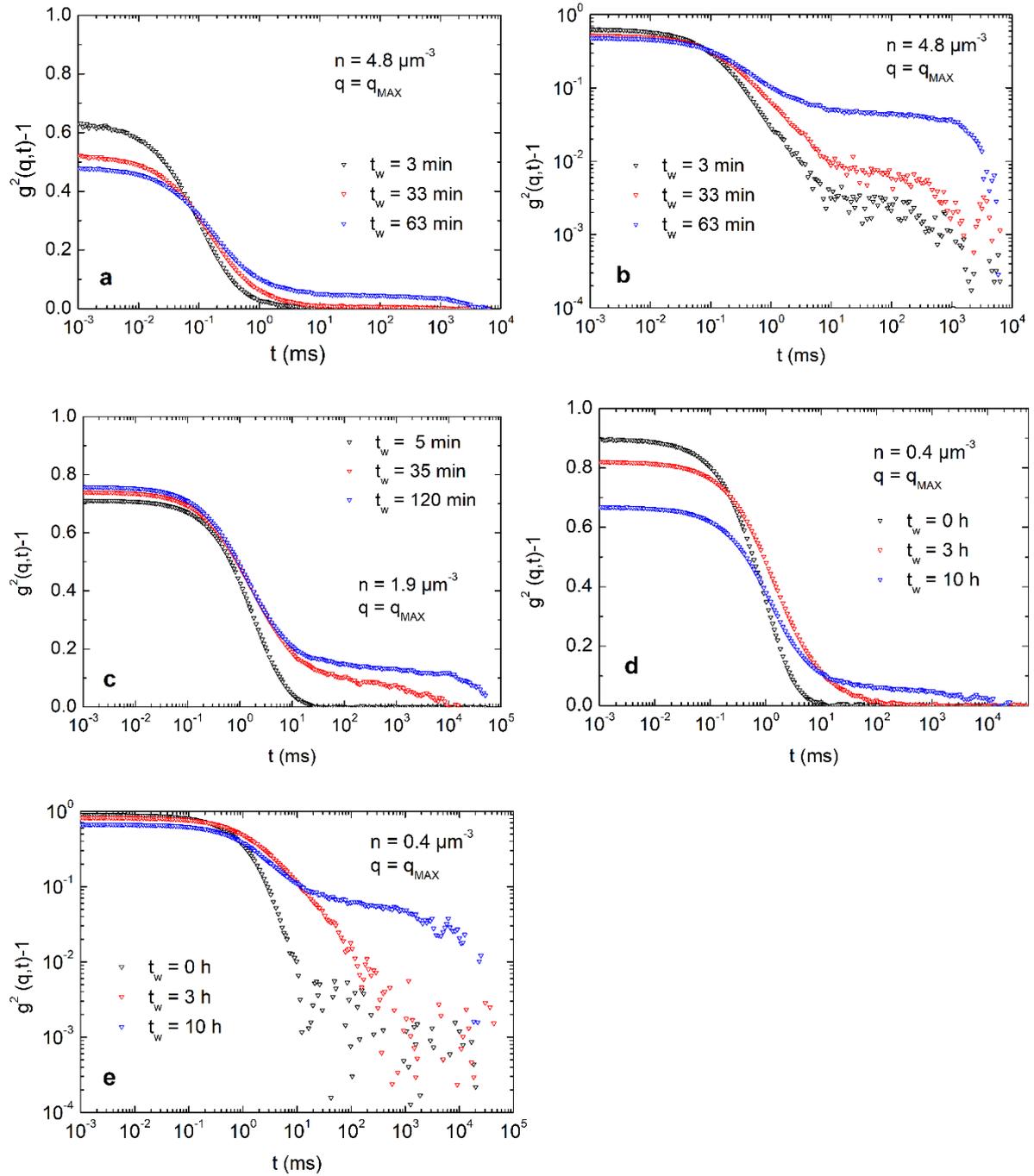

**Figure S5. Normalized intensity autocorrelation functions $g^2(q,t)-1$.** Data are plotted in both linear-logarithmic fashion (a, c and d) and double-logarithmic fashion (b and e) for vitrifying samples at different number densities of $n = 4.8$ μm$^{-3}$ ($\Phi = 3.8 \times 10^{-3}$), $n = 1.9$ μm$^{-3}$ ($\Phi = 1.3 \times 10^{-3}$) and $n = 0.4$ μm$^{-3}$ ($\Phi = 2.7 \times 10^{-4}$) as indicated. Measurements are shown for different waiting times. All measurements were performed at the respective $q_{MAX}$ of $q_{MAX} = 11.77$ μm$^{-1}$, 8.46 μm$^{-1}$ and 5.39 μm$^{-1}$, respectively.